\documentclass[12pt]{article}
\usepackage{a4wide}
\usepackage{amssymb}
\usepackage{graphicx}
\begin{document}
{\renewcommand{\thefootnote}{\fnsymbol{footnote}}
\hfill  CGPG--02/6--1\\
\medskip
\hfill gr--qc/0206053\\
\medskip
\begin{center}
{\LARGE Quantization Ambiguities in\\[2mm] Isotropic Quantum Geometry}\\
\vspace{1.5em}
Martin Bojowald\footnote{e-mail address: {\tt bojowald@gravity.psu.edu}}\\
\vspace{0.5em}
Center for Gravitational Physics and Geometry,\\
The Pennsylvania State
University,\\
104 Davey Lab, University Park, PA 16802, USA\\
\vspace{1.5em}
\end{center}
}

\setcounter{footnote}{0}

\newcommand{\case}[2]{{\textstyle \frac{#1}{#2}}}
\newcommand{\lP}{l_{\mathrm P}}
\newcommand{\md}{{\mathrm{d}}}
\newcommand{\tr}{\mathop{\mathrm{tr}}}
\newcommand{\sgn}{\mathop{\mathrm{sgn}}}

\newcommand*{\R}{{\mathbb R}}
\newcommand*{\Z}{{\mathbb Z}}

\begin{abstract}
 Some typical quantization ambiguities of quantum geometry are studied
 within isotropic models. Since this allows explicit computations of
 operators and their spectra, one can investigate the effects of
 ambiguities in a quantitative manner. It is shown that those
 ambiguities do not affect the fate of the classical singularity,
 demonstrating that the absence of a singularity in loop quantum
 cosmology is a robust implication of the general quantization
 scheme. The calculations also allow conclusions about modified
 operators in the full theory. In particular, using holonomies in a
 non-fundamental representation of $SU(2)$ to quantize connection
 components turns out to lead to significant corrections to classical
 behavior at macroscopic volume for large values of the spin of the
 chosen representation.
\end{abstract}

\section{Introduction}

Quantization consists in turning functions on the phase space of a
given classical system into operators acting on a Hilbert space
associated with the quantized system. To construct this map, one
selects a set of `elementary' observables (like $(q,p)$ in quantum
mechanics) which generate all functions on the phase space and form a
subalgebra of the classical Poisson algebra. This subalgebra has to be
mapped homomorphically into the quantum operator algebra, turning real
observables into selfadjoint operators (i.e., one is looking for a
$*$-representation of a suitable subalgebra of the classical Poisson
algebra on the quantum Hilbert space). Such a procedure is provided by
general quantization schemes (most relevant for quantum geometry is
algebraic quantization \cite{AlgQuant,ALMMT}, others are geometric
quantization \cite{Woodhouse} or group theoretic quantization
\cite{Isham}; in this context, see also \cite{Group} for a discussion
of allowed representations in the last scheme), but the issue of more
complicated, composite expressions is left open. In general, composite
operators are built up from the elementary ones in a way following
the classical expression. However, at this point quantization
ambiguities arise which render a unique quantization impossible.

The most obvious problem is the following one: Since the symplectic
structure on the classical phase space is non-degenerate, the operator
algebra must be non-abelian which leads to factor ordering ambiguities
for composite operators. More generally, the origin of quantization
ambiguities is that classical relations need not hold exactly after
quantization (in fact, not all classical relations can be exact at the
quantum level; an exact correspondence is required only for the
algebraic and reality conditions mentioned above). Therefore, we can
rewrite a given classical expression using some classical identity and
quantize the original and the rewritten expression; if the classical
identity is violated at the quantum level, the result will be two
different operators for the same classical expression. One can only
require that both operators coincide in some (classical) limit
involving $\hbar\to0$, but in a regime where quantum effects are
important they can be quite different from each other. Those
ambiguities can be fixed only by comparing with experimental
observations or, in the absence of experiments, by internal
consistency. If qualitative predictions involving quantum effects are
done, they need to be checked for independence of possible
ambiguities.

In many systems, quantization ambiguities can be largely ignored since
interesting observables are relatively simple. For instance, a
standard Hamiltonian in quantum mechanics of the form
$\frac{1}{2m}p^2+V(q)$ is free of factor ordering ambiguities. Another
example is that of generators of symmetries which have a distinguished
representation for the system in question, like the Gauss constraint in
Yang--Mills theories. In canonical quantum gravity (see
\cite{ThomasRev} for a recent review), however, the situation is
different: its dynamics is governed by the Hamiltonian constraint
which is a complicated expression and does not generate a simple
symmetry of functions on the space manifold. For a long time the
problem of finding even one well-defined quantization had remained out
of reach; only recently has an operator been given in quantum geometry
\cite{AnoFree,QSDI}. This was made possible by using classical
reformulations of some components of the classical constraint which
brought the expression into a form suitable for a quantization. As
discussed above, those reformulations also open the door for
quantization ambiguities and their effects have to be studied
carefully.

In quantum geometry, basic variables are holonomies along curves in
space associated with the Ashtekar connection $A_a^i$, and the
densitized triad $E^a_i$. Whereas the area of a surface or the volume
of a region in space can be written as functionals solely of the
densitized triad, and so have fairly unambiguous quantizations
\cite{AreaVol,Area,Vol2}, other geometric operators and also the
Hamiltonian constraint contain the co-triad
$e_a^i=|\det(E^b_j)|^{\frac{1}{2}}(E^{-1})^i_a$ which cannot be
quantized directly. Instead, the classical identity (with the
gravitational constant $\kappa=8\pi G$, the Barbero--Immirzi parameter
$\gamma$ \cite{Immirzi}, and the volume $V=\int\md^3x\sqrt{\det q}=
\int\md^3x|\det(E^b_j)|^{\frac{1}{2}}$)
\begin{equation}\label{cotriad}
 e_a^i=2(\gamma\kappa)^{-1}\{A_a^i,V\}
\end{equation}
is used \cite{QSDI} which can be quantized by using the known volume
operator, expressing the connection via holonomies, and turning the
Poisson bracket into a commutator. It is even possible to turn
singular classical expressions containing inverse powers of
$\sqrt{\det q}$ into well-defined operators because inverse powers of
the volume can be absorbed into the volume appearing in the Poisson
bracket. This is the mechanism which leads to well-defined, finite
matter Hamiltonians \cite{QSDV} and also plays an important role for
the absence of cosmological singularities \cite{Sing,InvScale}.

Using classical reformulations of this kind gives rise to quantization
ambiguities. Classically we have the relation
\begin{equation}\label{triadrel}
|\det (e_a^i)|=\sqrt{|\det(E^b_j)|}
\end{equation}
where the left hand side can be quantized using the expression
(\ref{cotriad}) of $e_a^i$ as a Poisson bracket, and the right hand
side using the basic $E^b_j$ (which has been done to derive the
standard volume operator \cite{Vol2}) resulting in very different
operators. More generally, we can insert
\begin{equation}\label{ambig1}
1=\left(\frac{|\det(e^i_a)|}{\sqrt{|\det(E_j^b)|}}\right)^k\,,
\end{equation}
which holds for any non-degenerate triad, into an arbitrary classical
expression and obtain a different quantization. (If we want to absorb
negative powers of $\sqrt{\det q}$ into the Poisson brackets to obtain
a well-defined operator, the power $k$ should be chosen to be
positive, but is arbitrary otherwise.) This is the first type of
ambiguity studied in this paper.

There is a second, quite distinct ambiguity which appears in the same
expression $\{A_a^i,V\}$ for $e_a^i$. For the quantization of $V$ we
have a unique choice, but as mentioned the connection components have
to be expressed via holonomies since those are basic in quantum
geometry. Initially, only holonomies in the fundamental representation
of $SU(2)$ have been used, but as first exploited in \cite{Gaul}, one
can choose an arbitrary representation as long as one compensates by a
constant prefactor (which just depends on the spin of the
representation). This constitutes a second one-parameter family of
ambiguities.

In the full theory, operators reflecting these ambiguities are
difficult to study explicitly since, e.g., the volume operator which
plays a prominent role in this kind of quantization is complicated and
its spectrum is not known explicitly. Therefore, here we study those
ambiguities within isotropic quantum geometry \cite{IsoCosmo}, where
the complete volume spectrum has been derived explicitly
\cite{cosmoII}. In this context we can compute all relevant operators
and their spectra and decide whether or not there are internal
constraints on the ambiguities. One result which has implications for
the full theory is that the spin chosen for the representation of
holonomies cannot be too high if almost classical behavior at
reasonably large volumes is to be preserved.

The considerations are also of interest within cosmological models:
for a singularity-free evolution \cite{Sing,IsoCosmo} it is necessary
that operators for the inverse scale factor \cite{InvScale} or inverse
volume annihilate the state $|0\rangle$ which corresponds to the
classical singularity. Since this state is also annihilated by the
volume operator, a simultaneous annihilation by the inverse volume
cannot correspond to classical behavior and so could be expected to be
sensitive to quantization ambiguities. In particular, the eigenvalue
in $|0\rangle$ must be zero, not just finite as expected if the
curvature divergence is regularized in the quantum theory. In
\cite{InvScale} it has been shown that the eigenvalue is in fact zero
for the particular quantization discussed there, and the following
general explanation has been presented. The procedure sketched above
leads to expressions where $\sqrt{\det q}=|\det e^i_a|$ appears within
a Poisson bracket. Since a Poisson bracket acts as a derivative on its
two arguments, there will always be factors of $\sgn(\det e^i_a)$
(which only matter if one looks at the behavior at the classical
singularity). If the rest of the expression is expected to have a
finite quantization in $|0\rangle$, it will be pushed to zero by the
sign factor. (There is no general proof, however, since the classical
expression involves the undefined quantity ``$0/0$''). Still, since
this result is crucial for a singularity-free evolution, it has to be
checked against possible quantization ambiguities. The quantization of
the inverse scale factor and related operators makes use of the
reformulation of $e_a^i$ and so we can use the ambiguities, discussed
above, to shed some light on this issue. It turns out that all
operators in the families of ambiguities studied here annihilate the
state $|0\rangle$, and so the predictions of loop quantum cosmology
concerning the fate of the classical singularity are robust.

The plan of the paper is as follows: In the following section we
recall the basic formulas of isotropic loop quantum cosmology, mainly
the action of operators which will be used later. Using this framework,
a quantization of the inverse volume will be introduced in Section
\ref{AmTriad} which realizes the first quantization ambiguity,
followed in Section \ref{AmHol} by a quantization of the inverse
square root of the scale factor realizing the second ambiguity. The
choice of these particular powers of the scale factor is just to
obtain simple quantizations which can be studied easily. For other
expressions the same steps can be repeated and we can also combine
both families of ambiguities in one operator. Finally, consequences
regarding the full theory and quantum cosmology are discussed in
Section~\ref{Concl}.

\section{Basic Operators}

To find symmetric states one needs to know the general form of
connections and triads invariant under the given action of the
symmetry group on space \cite{PhD,SymmRed}. In the case of homogeneity
\cite{cosmoI}, invariant connections and triads have the form
$A_a^i=\phi_I^i\omega_a^I$, $E^a_i=p^I_i X^a_I$, where $\omega^I_a$
and $X^a_I$ are dual one-forms and vector fields on space which are
left-invariant under the symmetry group. For homogeneity, the
coefficients $\phi_I^i$, $p^I_i$ are arbitrary constants, but for
isotropic models they must have the form $\phi^i_I=c\Lambda^i_I$,
$p^I_i=p\Lambda^I_i$ where $\Lambda^I_i$ is an internal $SU(2)$-triad
which is rotated under gauge transformations, and $c$ and $p$ are the
only gauge invariant parameters with the symplectic structure
$\{c,p\}=\case{1}{3}\gamma\kappa$ (using the Barbero--Immirzi
parameter $\gamma$). The isotropic triad component $p$ can be any real
number; its sign determines the orientation of the triad. It is
related to the scale factor $a$, which appears as the isotropic
co-triad component $e_a^i=a^i_I\omega^I_a=a\Lambda_I^i\omega^I_a$,
through $p=\sgn(a)a^2$ (note that we also have a sign in $a$ due to
the two possible orientations).

In the connection representation, gauge invariant isotropic states are
represented as function of the single parameter $c$; an orthonormal
basis is given by \cite{cosmoII,InvScale}
\begin{equation}\label{n}
 |n\rangle:=\frac{\exp(\case{1}{2}inc)}{\sqrt{2}\sin\case{1}{2}c}
 \quad,\quad n\in\Z\,.
\end{equation}
Basic multiplication operators are given by the three holonomies $h_I:=
\exp(c\Lambda_I^i\tau_i)= \cos(\frac{1}{2}c)+ 2\sin(\frac{1}{2}c)
\Lambda_I^i\tau_i$ which are all gauge rotations of each other. The
action of operators containing these holonomies can always be computed
using the basic relations
\begin{eqnarray}
 \cos(\case{1}{2}c) |n\rangle &=& \case{1}{2}(|n+1\rangle+|n-1\rangle)\\
 \sin(\case{1}{2}c) |n\rangle &=& -\case{1}{2}i(|n+1\rangle-
 |n-1\rangle)\,.
\end{eqnarray}
For geometrical operators we also need the volume operator whose
action is \cite{cosmoII,InvScale}
\begin{equation} \label{vol}
 \hat{V}|n\rangle= V_{\frac{1}{2}(|n|-1)}|n\rangle=
 (\case{1}{6}\gamma\lP^2)^{\frac{3}{2}} \sqrt{(|n|-1)|n|(|n|+1)}
   \,|n\rangle\,.
\end{equation}
It will appear mainly in the form of a commutator with a holonomy:
\begin{eqnarray} \label{hVcomm}
 h_I[h_I^{-1},\hat{V}] &=& \hat{V}-\cos(\case{1}{2}c) \hat{V}
 \cos(\case{1}{2}c)- \sin(\case{1}{2}c) \hat{V} \sin(\case{1}{2}c)\\
 &&- 2\Lambda_I^i\tau_i \left(\sin(\case{1}{2}c) \hat{V} \cos(\case{1}{2}c)-
 \cos(\case{1}{2}c) \hat{V} \sin(\case{1}{2}c)\right)\,. \nonumber
\end{eqnarray}
The second term of this sum has the action
\begin{eqnarray}\label{sVc}
 \left(\sin(\case{1}{2}c) \hat{V}
   \cos(\case{1}{2}c)-\cos(\case{1}{2}c) \hat{V}
   \sin(\case{1}{2}c)\right)|n\rangle &=& \case{1}{2}i \left(
   V_{\frac{1}{2}(|n+1|-1)}- V_{\frac{1}{2}(|n-1|-1)}\right) |n\rangle
 \nonumber\\
 &=& \case{1}{2}i \sgn(n)
 \left(V_{\frac{1}{2}|n|}-V_{\frac{1}{2}|n|-1}\right) |n\rangle
\end{eqnarray}
which directly shows its eigenvalues.

An example for an operator containing these expressions is the inverse
scale factor operator of \cite{InvScale}. One starts by writing
$|a|^{-1}\delta_{IJ}= |\det (a^k_K)|^{-1} a^i_Ia^i_J$ in terms of
isotropic co-triad components and uses the identity (\ref{cotriad})
for $a^i_I$. The quantization then proceeds by using the volume
operator, writing the connection components in terms of holonomies and
turning the Poisson bracket into a commutator. Important for the
finiteness of the resulting operator is that the determinant of
$a^i_I$, by which we have to divide in the classical expression, can
be absorbed into the volume appearing in the Poisson bracket. Thus,
the inverse scale factor has a well-defined quantization even though
the volume operator has eigenvalue zero. This procedure is a common
technique in quantum geometry \cite{QSDI,QSDV}, which uses
(\ref{cotriad}) as an essential ingredient. Since it applies to a
large number of interesting quantities which are not simple functions
of the basic operators, it is not surprising that one has to
reformulate the original classical expression, and quantization
ambiguities are to be expected.

\section{Ambiguities in Triad Quantizations}
\label{AmTriad}

We start with the first kind of ambiguities mentioned in the
Introduction, namely the one resulting from a possible violation of
the classical relation $|\det(e_a^i)|=|\det(E^b_j)|^{\frac{1}{2}}$ at
the quantum level. For illustrative purposes we discuss an operator
whose absolute value on non-degenerate states is a quantization of the
inverse volume $V^{-1}=|\det(e_a^i)|^{-1}$. The volume operator
(\ref{vol}) has eigenvalue zero, and so its inverse does not exist as
a densely defined operator. However, as in \cite{InvScale} we can use
the techniques developed in \cite{QSDI} to quantize the inverse
volume; in fact this has already been done for matter Hamiltonians in
\cite{QSDV} where the first step is to replace $(\det
q)^{-\frac{1}{2}}$ by $\det (e_a^i)^2 (\det q)^{-\frac{3}{2}}$. Here
(\ref{triadrel}) has been used, and this reformulation allows to apply
(\ref{cotriad}) and a subsequent quantization to a well-defined
operator. We are going to follow the same procedure in the reduced
model of isotropic quantum geometry, but allowing arbitrary powers as
in (\ref{ambig1}). This leads us to
\begin{eqnarray} \label{rkdef}
 r_k:=\frac{\sgn(\det(a^i_I))^k}{\sqrt{\det q}} &=&
 \left(\frac{\det(a^i_I)}{\sqrt{\det q}^{1+k^{-1}}}\right)^k \\
 &=& \left(\case{4}{3}(\kappa\gamma)^{-3} \epsilon^{IJK}\epsilon_{ijk}
 \frac{\{\phi_I^i,V\} \{\phi^j_J,V\}
 \{\phi_K^k,V\}}{V^{1+k^{-1}}}\right)^k \nonumber
\end{eqnarray}
where we used the relation $a_I^i=2(\kappa\gamma)^{-1} \{\phi_I^i,V\}$
for the homogeneous co-triad which is the analog of (\ref{cotriad})
(see \cite{InvScale} for a derivation which takes care of the sign of
$\det(a_I^i)$). In the next step we absorb the volume in the
denominator into the $V$ appearing in the Poisson brackets and write
the contraction of the coefficients $\phi_I^i$ in the $SU(2)$-indices
as a trace over a product of $\phi_I:=\phi_I^i\tau_i$:
\[
 r_k= \left(-\case{16}{3}(\kappa\gamma)^{-3}
 \!\left(\!\frac{3}{2-k^{-1}}\right)^3 \!\epsilon^{IJK} \tr\left(
 \{\phi_I,V^{\frac{1}{3}(2-k^{-1})}\}
 \{\phi_J,V^{\frac{1}{3}(2-k^{-1})}\}
 \{\phi_K,V^{\frac{1}{3}(2-k^{-1})}\} \right)\right)^k\,.
\]
Note that we distributed the volume in the denominator symmetrically
over the three Poisson brackets; choosing a different way gives rise
to another ambiguity which is briefly discussed below. The last
expression can be quantized immediately by turning
$\{\phi_I,V^{\frac{1}{3}(2-k^{-1})}\}$ into $i\hbar^{-1}
h_I [h_I^{-1},\hat{V}^{\frac{1}{3}(2-k^{-1})}]$, resulting in
\begin{eqnarray}
 \hat{r}_k \!\!\!\!&=&\!\!\!\!\! \left(\! 144i
 ((2\!-\!k^{-1})\gamma\lP^2)^{-3} \epsilon^{IJK}\! \tr\!\left(
 h_I[h_I^{-1},\hat{V}^{\frac{1}{3}(2-k^{-1})}]
 h_J[h_J^{-1},\hat{V}^{\frac{1}{3}(2-k^{-1})}]
 h_K[h_K^{-1},\hat{V}^{\frac{1}{3}(2-k^{-1})}] \right)\!\!\right)^k
 \nonumber\\ 
 \!\!\!\!&=&\!\!\!\!\! \left(\!
 i(\gamma\lP^2)^{-3} \left( \frac{12}{2-k^{-1}} \right)^3 
 \left(\sin(\case{c}{2}) \hat{V}^{\frac{1}{3}(2-k^{-1})}
 \cos(\case{c}{2})- \cos(\case{c}{2}) \hat{V}^{\frac{1}{3}(2-k^{-1})}
 \sin(\case{c}{2}) \right)^3\right)^k
\end{eqnarray}
using (\ref{hVcomm}) and $\tr(\Lambda_I^i\tau_i \Lambda_J^j\tau_j
\Lambda_K^k\tau_k)= -\case{1}{4}\epsilon_{IJK}$.  Such an operator is
densely defined as long as we choose $k>\frac{1}{2}$ since in this
case we only have positive powers of the volume operator. This
quantization corresponds to the one in \cite{InvScale} for $k=2$ with
the cubic root of $\hat{r}_2$ giving the inverse scale factor
operator.

\begin{figure}[ht]
\begin{center}
 \includegraphics[width=12cm,height=8cm,keepaspectratio]{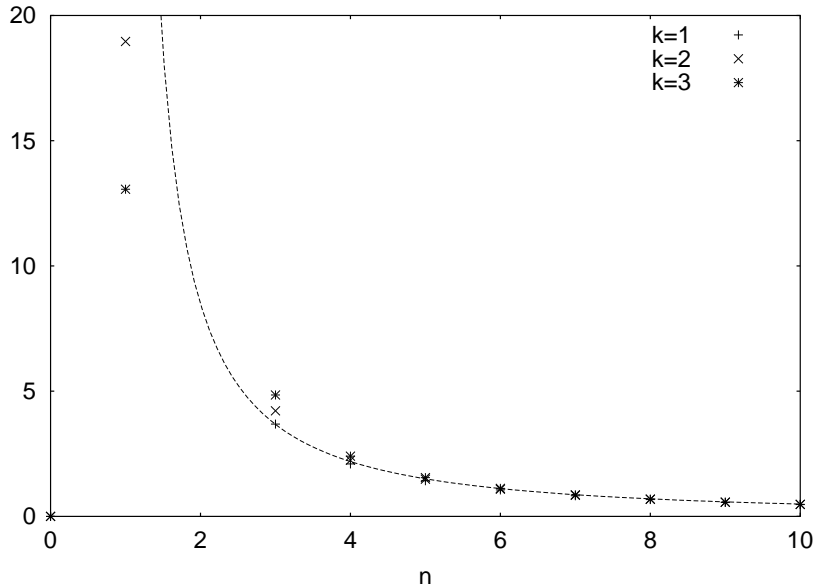}
\end{center}
\caption{Eigenvalues of the inverse volume (in units of
$(\gamma\lP^2)^{-\frac{3}{2}}$) compared to the inverse eigenvalues of
the volume operator (dashed line). The eigenvalues for $n=0$ are all
zero, independently of $k$, and so lie on top of each other in this
diagram. For $n=2$ and $n=1$, $k=1$, the eigenvalues are cut off and
can be seen in Fig.~\ref{Vinvlarge}. The inverse scale factor operator
of \cite{InvScale} corresponds to the cubic root of the values for
$k=2$.}
\label{Vinv}
\end{figure}

\begin{figure}[ht]
\begin{center}
 \includegraphics[width=12cm,height=8cm,keepaspectratio]{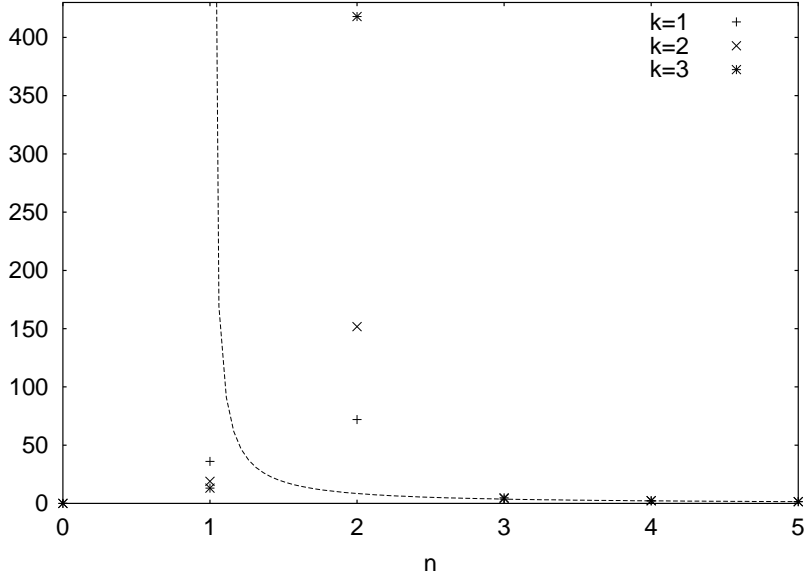}
\end{center}
\caption{The lowest eigenvalues of the inverse volume (in units of
$(\gamma\lP^2)^{-\frac{3}{2}}$) compared to the inverse eigenvalues of
the volume operator (dashed line).}
\label{Vinvlarge}
\end{figure}

The eigenvalues of this family of operators are readily determined
using the basic operators $\sin(\case{c}{2})$, $\cos(\case{c}{2})$ and
$\hat{V}$ (or directly (\ref{sVc})):
\begin{equation}\label{rkn}
 r_{k,n}=(\gamma\lP^2)^{-3k} 
 \left(\frac{6\sgn n}{2-k^{-1}}\right)^{3k}
 \left(V_{\frac{1}{2}|n|}^{\frac{1}{3}(2-k^{-1})}-
 V_{\frac{1}{2}|n|-1}^{\frac{1}{3}(2-k^{-1})}\right)^{3k}\,.
\end{equation}
They are bounded from above and vanish for $n=0$, a property
which is important for the absence of a singularity. Fig.~\ref{Vinv}
shows the eigenvalues for $k=1$, $k=2$, $k=3$ (the values for $n=2$
and the $n=1$-value for $k=1$ are cut off in Fig.~\ref{Vinv}, but can
be seen in Fig.~\ref{Vinvlarge}).

We have the largest deviations between the classical expectation and
the eigenvalues for very small volume ($n<3$), where also the
strongest dependence on the ambiguity labelled by $k$ occurs; see
Fig.~\ref{Vinvlarge}. For $n=1$, where the volume eigenvalue is zero,
we have the finite eigenvalue $r_{k,1}=6(\gamma\lP^2)^{-\frac{3}{2}}
(1-(2k)^{-1})^{-3k}\cdot (\case{3}{4})^k$ of the inverse volume which
decreases with $k$. For $n=2$ the inverse volume has eigenvalue
$r_{k,2}=3(\gamma\lP^2)^{-\frac{3}{2}} (1-(2k)^{-1})^{-3k}\cdot 3^k$
which increases exponentially for $k>1$ and has a minimum at $k_{\rm
min}\simeq 0.95$ (an exact expression is $k_{\rm min}= \frac{1}{2}
W(-(3^{\frac{1}{3}}e)^{-1})/ (1+W(-(3^{\frac{1}{3}}e)^{-1}))$ in terms
of the Lambert function $W(x)$ which satisfies $W(x)\exp W(x)=x$) with
value $r_{k_{\rm min},2}\simeq72(\gamma\lP^2)^{-\frac{3}{2}}$. The
value at $n=2$ determines the upper bound for the inverse volume
eigenvalues (which can be interpreted as an upper bound for curvature)
and so is smallest if we choose the quantization with $k=1$. (We could
also allow non-integer values for $k$ if we work only with the
absolute value of $r_k$, but choosing the true minimum $k_{\rm min}$
would not change the upper bound significantly.) So the mimimal
allowed integer value $k=1$ gives smallest curvature and results which
are closest to the inverse volume eigenvalues at small values of $n$.

\begin{figure}[ht]
\begin{center}
 \includegraphics[width=12cm,height=8cm,keepaspectratio]{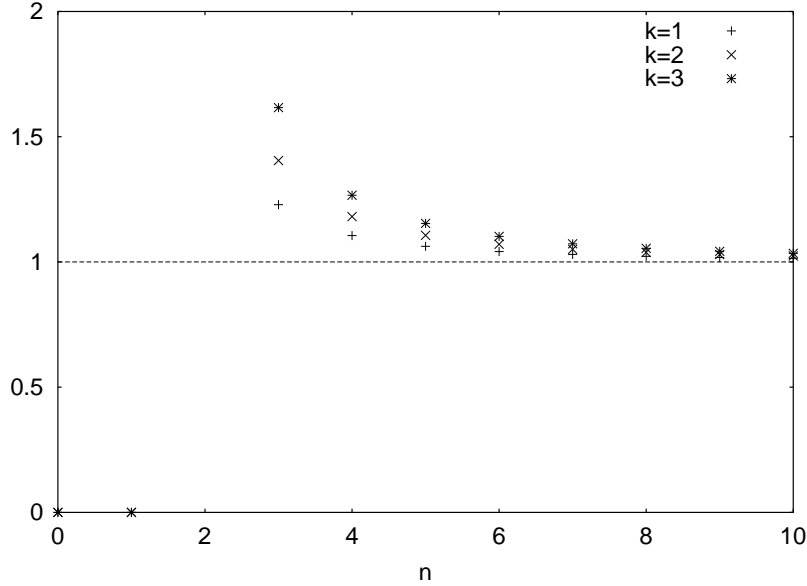}
\end{center}
\caption{Product of the eigenvalues of the inverse volume with the
volume eigenvalues compared to the classical expectation one (dashed
line). For $n=0$ and $n=1$, the products of eigenvalues are zero
independently of $k$; the eigenvalues for $n=2$ are cut off.}
\label{Vprod}
\end{figure}

If $|n|$ is large (which means greater than $2$) we also observe
that $k=1$ gives eigenvalues for $r_k$ which are closest to the
inverse volume eigenvalues. For all shown values the approach to the
classical expectation is rapid and already starts at volumes which are
not necessarily large compared to the Planck scale; this can be seen
in Figs.~\ref{Vinv} and \ref{Vprod}.
If we assume $|n|\gg1$, we have the expansion
\begin{eqnarray}\label{rkexpand}
 |r_{k,n}| &=& V_{\frac{1}{2}(|n|-1)}^{-1} \left(1+(k+\case{1}{4}+
 \case{1}{8}k^{-1}) n^{-2} +O(n^{-4})\right) \nonumber\\
 &\sim& V^{-1} \left(1+\case{1}{36}(k+\case{1}{4}+\case{1}{8}k^{-1})
 \gamma^2\lP^4 a^{-4}\right)
\end{eqnarray}
using $a^2\sim\frac{1}{6}\gamma\lP^2|n|$ at large $|n|$ in the last
step, which follows from the volume eigenvalues (\ref{vol}).  For
large $k$, the correction term increases linearly in $k$; it is
minimal for $k=\frac{1}{2\sqrt{2}}$, so also in this regime $k=1$ is
the integer for which the behavior is closest the the classical one.

To close this section we briefly discuss the ambiguity arising from a
non-symmetric distribution of the inverse volume over the three
Poisson brackets. This ambiguity is a two-parameter family which we
describe by introducing three new parameters $x$, $y$, $z$ subject to
the condition $x+y+z=1$. Instead of the original $r_k$ we then have
\begin{eqnarray*}
 r_{k,x,y,z} &:=&\left(\case{4}{3}(\kappa\gamma)^{-3}
 \epsilon^{IJK}\epsilon_{ijk}
 \frac{\{\phi_I^i,V\} \{\phi^j_J,V\}
 \{\phi_K^k,V\}}{V^{x(1+k^{-1})}V^{y(1+k^{-1})}V^{z(1+k^{-1})}}\right)^k\\
 &=& \left(-\case{16}{3}(\kappa\gamma)^{-3}
 \epsilon^{IJK} \frac{\tr\left(
 \{\phi_I,V^{1-x(1+k^{-1})}\}
 \{\phi_J,V^{1-y(1+k^{-1})}\}
 \{\phi_K,V^{1-z(1+k^{-1})}\} \right)}{ -k^{-1}+
 (xy+xz+yz)(1+k^{-1})^2- xyz(1+k^{-1})^3} \right)^k
\end{eqnarray*}
which after quantizing leads to an operator with eigenvalues
\begin{eqnarray*}
 r_{k,x,y,z,n} &=&
 (2\sgn n)^{3k} (\gamma\lP^2)^{-3k} (-k^{-1}+
 (xy+xz+yz)(1+k^{-1})^2- xyz(1+k^{-1})^3)^{-k}\\
 && \times
 \left(V_{\frac{1}{2}|n|}^{1-x(1+k^{-1})}-
 V_{\frac{1}{2}|n|-1}^{1-x(1+k^{-1})}\right)^k
 \left(V_{\frac{1}{2}|n|}^{1-y(1+k^{-1})}-
 V_{\frac{1}{2}|n|-1}^{1-y(1+k^{-1})}\right)^k\\
 &&\times
 \left(V_{\frac{1}{2}|n|}^{1-z(1+k^{-1})}-
 V_{\frac{1}{2}|n|-1}^{1-z(1+k^{-1})}\right)^k\,.
\end{eqnarray*}
For $x=y=z=\frac{1}{3}$ this formula reduces to the symmetric case
which we already obtained in (\ref{rkn}).

The procedure leads to a well-defined operator if and only if all
powers of the volume operator are positive, i.e.\ $1-x(1+k^{-1})>0$
and similarly for $y$, $z$ which implies $x,y,z<k/(1+k)$ (so a single
parameter can be negative, as long as we satisfy the condition
$x+y+z=1$; e.g., $x=-\frac{1}{2}$, $y=z=\frac{3}{4}$ for $k\geq4$ for
which we have huge values
$r_{4,-\frac{1}{2},\frac{3}{4},\frac{3}{4},1}\sim10^6$,
$r_{4,-\frac{1}{2},\frac{3}{4},\frac{3}{4},2}\sim10^9$). The behavior
at large $n$ is not significantly altered if we choose $x$, $y$, and
$z$ different from $\frac{1}{3}$; in fact, we have
\begin{eqnarray*}
 |r_{k,x,y,z,n}| &=& V_{\frac{1}{2}(|n|-1)}^{-1}
 \left(1+\case{1}{8}k(7+3(x^2+y^2+z^2)(1+k^{-1})^2)n^{-2}+
 O(n^{-4})\right)\\
 &\sim& V^{-1}\left(1+\case{1}{288}k
 (7+3(x^2+y^2+z^2)(1+k^{-1})^2) \gamma^2\lP^4a^{-4}\right)\,.
\end{eqnarray*}
The correction is minimal for $k$ small and $x=y=z=\frac{1}{3}$ (which
minimizes $x^2+y^2+z^2$ under the condition $x+y+z=1$). Furthermore,
for $n=0$ we always have eigenvalue zero; but the values for $n=1$ and
$n=2$ do depend strongly on $x$, $y$, $z$. It turns out that both are
minimal in the symmetric case $x=y=z=\frac{1}{3}$ (independently of
$k$), so also this ambiguity leads to the smallest curvature and to
smallest deviations from classical behavior for the simplest
quantization.

\section{Representation of Holonomies}
\label{AmHol}

We now turn to the second class of ambiguities studied in this
paper. As noticed in \cite{Gaul}, there is a freedom in choosing
representations for holonomies appearing in an operator. Usually, the
fundamental representation is used for simplicity, but one could be
led to other representations by enforcing certain properties. Here we
study a simple operator which contains only one commutator of a
holonomy with the volume operator. Classically, this is the inverse
square root of the scale factor $a$ which we write as
\begin{equation}\label{sjdef}
 s_j:=\frac{\sgn(a)}{\sqrt{|a|}}=\frac{a}{\sqrt{V}}=\frac{\Lambda^I_i
 a^i_I}{3\sqrt{V}}= -\frac{\tr(\Lambda^{Ik}\tau_k^{(j)}
 a_I^i\tau_i^{(j)})}{j(j+1)(2j+1)\sqrt{V}}=
 -\frac{4\tr(\Lambda^{Ik}\tau_k^{(j)} \{\phi_I^i,\sqrt{V}\}
 \tau_i^{(j)})}{j(j+1)(2j+1)\gamma\kappa}
\end{equation}
using the formula
$\tr(\tau_i^{(j)}\tau_k^{(j)})=-\case{1}{3}j(j+1)(2j+1)\delta_{ik}$
for the trace in a representation $j$.

Following the usual steps, this will be quantized to
\begin{eqnarray}
 \hat{s}_j &=& -4i(j(j+1)(2j+1)\gamma\lP^2)^{-1} \sum_I
 \tr\left(\Lambda^{Ik}\tau_k^{(j)} h_I^{(j)}
 \left[(h_I^{(j)})^{-1},\hat{V}^{\frac{1}{2}}
 \right] \right)\nonumber\\ 
 &=& -12i(j(j+1)(2j+1)\gamma\lP^2)^{-1}
 \tr\left(\Lambda^{3k}\tau_k^{(j)} h_3^{(j)}
 \left[(h_3^{(j)})^{-1},\hat{V}^{\frac{1}{2}} \right]\right)\,.
\end{eqnarray}
To compute this operator explicitly we exploit the fact that it is
gauge invariant, and so we can apply a gauge transformation which
rotates $\Lambda^3_k\tau^k$ into $\tau_3$. In the $j$-representation
we then have
\begin{equation}\label{hmatrix}
 h_3^{(j)}=\exp(c\tau_3^{(j)})=\left(\begin{array}{ccccc} e^{-ijc} & 0 &
 \cdots & & 0\\ 0& e^{-i(j-1)c} &&&\\ \vdots && \ddots &&\vdots\\ &&&
 e^{i(j-1)c} & 0\\ 0 &&\cdots&0&e^{ijc} \end{array}\right)
\end{equation}
with matrix elements $h_{3,\alpha\beta}^{(j)}=
e^{i(\alpha-j)c}\delta_{(\alpha)\beta}$ for $0\leq\alpha,\beta\leq
2j$. For the commutator (where $\hat{V}$ can be replaced with any
operator which is diagonal in the $|n\rangle$-basis), we need
\begin{eqnarray*}
 h_{3,\alpha\beta}^{(j)}\left[(h_3^{(j)})^{-1}_{\beta\gamma},\hat{V}\right]
 &=& h_{3,\alpha\beta}^{(j)} \left[\overline{h}_{3,\gamma\beta}^{(j)},
 \hat{V}\right]= 
 \delta_{(\alpha)\gamma}e^{i(\alpha-j)c} [e^{-i(\alpha-j)c},\hat{V}]\\
 &=& \delta_{(\alpha)\gamma}(\hat{V}- e^{i(\alpha-j)c} \hat{V}
 e^{-i(\alpha-j)c})
\end{eqnarray*}
which acts
\begin{equation}\label{commj}
 h_{3,\alpha\beta}^{(j)}\left[(h_3^{(j)})^{-1}_{\beta\gamma},\hat{V}\right]
 |n\rangle= \delta_{(\alpha)\gamma} (V_{\frac{1}{2}(|n|-1)}-
 V_{\frac{1}{2}(|n-2\alpha+2j|-1)}) |n\rangle\,.
\end{equation}

If we multiply the commutator with
$\tau^{(j)}_{3,\alpha\beta}=i(\alpha-j)\delta_{(\alpha)\beta}$ and take
the trace, we obtain
\[
 \tr\left(\tau_3^{(j)}h_3^{(j)} \left[(h^{(j)}_3)^{-1},\hat{V}\right]
 \right)|n\rangle= i\sum_{k=-j}^j kV_{\frac{1}{2}(|n+2k|-1)} |n\rangle
\]
which directly gives the eigenvalues
\begin{equation}\label{sjn}
 s_{j,n}= 12 (j(j+1)(2j+1)\gamma\lP^2)^{-1}\sum_{k=-j}^j
 k\sqrt{V_{\frac{1}{2}(|n+2k|-1)}}
\end{equation}
for the inverse square root of the scale factor.

\begin{figure}[ht]
\begin{center}
 \includegraphics[width=12cm,height=8cm,keepaspectratio]{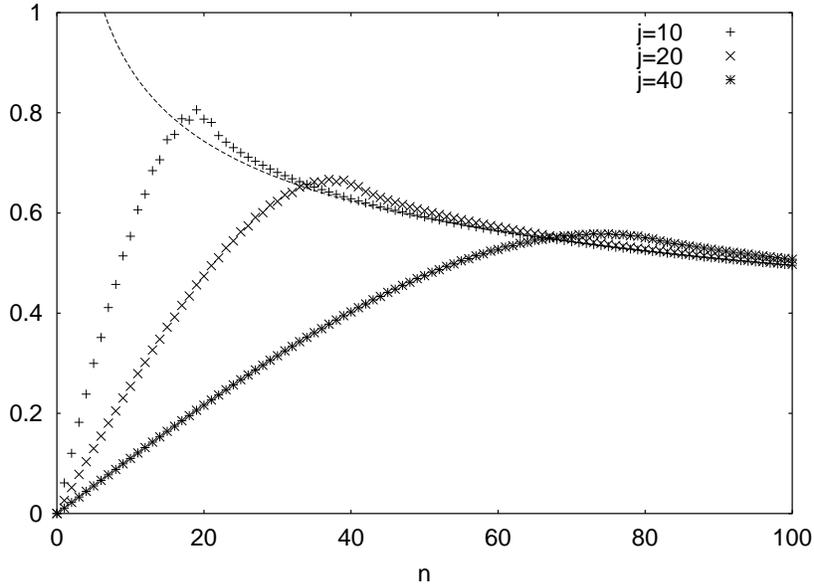}
\end{center}
\caption{Eigenvalues of the inverse square root of the scale factor
(in units of $(\gamma\lP^2)^{-\frac{1}{4}}$) compared to the
eigenvalues $V_{\frac{1}{2}(|n|-1)}^{-\frac{1}{6}}$ (dashed line).}
\label{Ainv}
\end{figure}

\begin{figure}[ht]
\begin{center}
 \includegraphics[width=12cm,height=8cm,keepaspectratio]{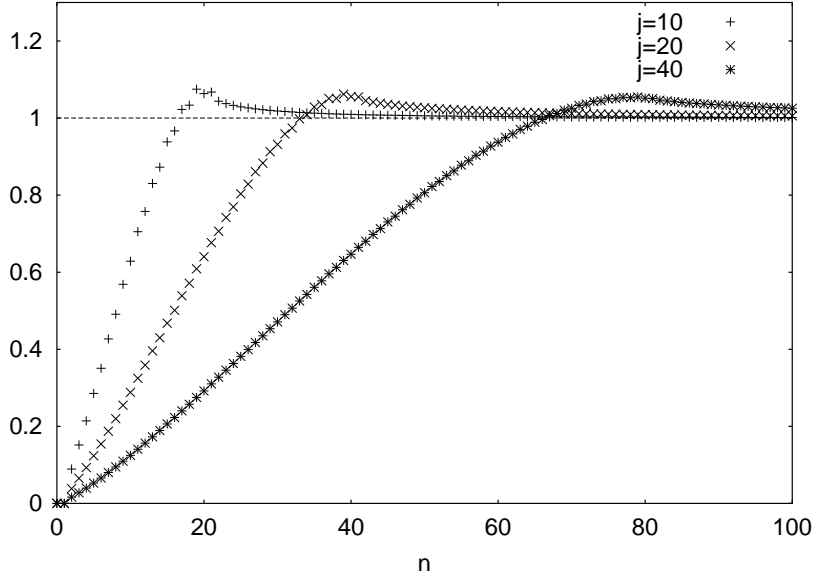}
\end{center}
\caption{Product of the eigenvalues of the inverse square root of the
scale factor with $V_{\frac{1}{2}(|n|-1)}^{\frac{1}{6}}$
compared to the classical expectation one (dashed line).}
\label{Aprod}
\end{figure}

Again, for all values of $j$ the eigenvalue in the state $|0\rangle$
vanishes, implying that the results about the fate of the classical
singularity are independent of this ambiguity. But here the effects of
changing $j$ are more dramatic for non-zero $n$: If we approach $n=0$
from large values, then the increase of $|a|^{-\frac{1}{2}}$, which can
be observed in the classical curve, stops around the value $|n|=2j$ for
the eigenvalues of $\hat{s}_j$ and turns into a curve decreasing
toward zero which is reached for $n=0$ (see Figs.~\ref{Ainv},
\ref{Aprod}). This implies that the upper curvature bound is lower
than in the original quantization with $j=\frac{1}{2}$, but it also
means that deviations from the classical behavior already set in at
larger volumes. In particular, there is an upper bound for the allowed
values of $j$ in this quantization ambiguity even though it would be
quite large.

For large $|n|$, we have
\[
 \sqrt{V_{\frac{1}{2}(|n+2k|-1)}}=
 (\case{1}{6}\gamma\lP^2|n|)^{\frac{3}{4}} \left(1+\case{3}{2}kn^{-1}-
 \case{1}{8} (3k^2+2)n^{-2}+ \case{5}{16} (k^3+2k)n^{-3}+ O(n^{-4})\right)
\]
which implies
\begin{eqnarray}
 |s_{j,n}|&=&2
 (j(j+1)(2j+1))^{-1}(\case{1}{6}\gamma\lP^2|n|)^{-\frac{1}{4}} 
 \left(\case{3}{2}\sum_{k=-j}^j k^2+ \case{5}{16} \sum_{k=-j}^j
 (k^4+2k^2)n^{-2} +O(n^{-4})\right)\nonumber\\
 &=& (\case{1}{6}\gamma\lP^2|n|)^{-\frac{1}{4}} \left(1+\case{1}{8}
 (j^2+j-3)n^{-2} +O(n^{-4})\right) \nonumber\\
 &=& V_{\frac{1}{2}(|n|-1)}^{-\frac{1}{6}} \left(1+\case{1}{8}
 (j^2+j-\case{11}{3}) n^{-2}+ O(n^{-4})\right)\nonumber\\
 &\sim& V^{-\frac{1}{6}} \left(1+ \case{1}{288} (j^2+j-\case{11}{3})
 \gamma^2\lP^4a^{-4}\right) \label{sjnexpand}
\end{eqnarray}
using $\sum_{k=1}^j k^2=\case{1}{6}j(j+1)(2j+1)$ and $\sum_{k=1}^jk^4=
\case{1}{30}j(j+1)(2j+1)(3j^2+3j-1)$ (if $j$ is a half-integer, we
need the sums $\sum_{k=1}^{j+\frac{1}{2}} (k-\frac{1}{2})^2$ and
$\sum_{k=1}^{j+\frac{1}{2}} (k-\frac{1}{2})^4$, which have the same
explicit expressions as the sums for integer $j$). The correction term
is smallest for small $j$; in fact, as already observed, it can be
significant even for large volume, provided only that $j$ is large
enough. The behavior is close to classical if the ratio $n/j$ is
large.

This observation suggests to compute $s_{j,2qj}$ where $q$ is a new
parameter (i.e., we measure $n$ in multiples of $2j$) under the
assumptions $j\gg1$ and $qj\gg1$. Because the absolute value of
$2qj-2k$ appears in the volume eigenvalues in formula (\ref{sjn}),
we have to split the calculation into two parts, first assuming $q\geq
1$ and then $q<1$ (note that our condition $qj\gg1$ can be fulfilled
also for small values of $q<1$ provided that $j$ is large enough).

If $q>1$, we can simply drop all absolute values and obtain
\[
 s_{j,2qj}=12(j(j+1)(2j+1)\gamma\lP^2)^{-1} \sum_{k=1}^j
 k\left(\sqrt{V_{\frac{1}{2}(2(qj+k)-1)}}-
 \sqrt{V_{\frac{1}{2}(2(qj-k)-1)}}\right)\,.
\]
Here we have $qj+k\gg1$ for all values of $k$, and for most values we
have also $qj-k\gg1$ so that we have only a small error if we use this
condition for all values of $k$. Together with $j\gg1$
we then get
\begin{eqnarray*}
 s_{j,2qj} &\simeq& 2j^{-3}(\case{1}{3}\gamma\lP^2)^{-\frac{1}{4}}
 \sum_{k=1}^j k\left((qj+k)^{\frac{3}{4}}- (qj-k)^{\frac{3}{4}}\right) \\
 &=& 2j^{-3}(\case{1}{3}\gamma\lP^2)^{-\frac{1}{4}} \sum_{k=1}^j \left(
 (qj+k)^{\frac{7}{4}}-qj(qj+k)^{\frac{3}{4}}+
 (qj-k)^{\frac{7}{4}}-qj(qj-k)^{\frac{3}{4}} \right)\\
 &=& 2j^{-3}(\case{1}{3}\gamma\lP^2)^{-\frac{1}{4}}
 \left(\sum_{l=qj+1}^{(q+1)j}
 (l^{\frac{7}{4}}-qjl^{\frac{3}{4}})+ \sum_{l=(q-1)j}^{qj-1}
 (l^{\frac{7}{4}}-qjl^{\frac{3}{4}})\right)\\
 &=& 2j^{-3}(\case{1}{3}\gamma\lP^2)^{-\frac{1}{4}} \left(\sum_{l=1}^{(q+1)j}
 (l^{\frac{7}{4}}-qjl^{\frac{3}{4}})-
 \sum_{l=1}^{(q-1)j-1}(l^{\frac{7}{4}}-qjl^{\frac{3}{4}})\right)
\end{eqnarray*}
where we wrote $k=qj+k-qj=-(qj-k-qj)$, and relabelled $l=qj+k$ and
$l=qj-k$, respectively. In the next step we use again that $qj$ is
large so that we are summing over a large range of values. Then the
sums can be approximated by integrals in the following way: we have
$\int_0^1 x^r\md x=\sum_{l=1}^N (l/N)^rN^{-1}(1+O(N^{-1}))$ which
yields $\sum_{l=1}^N l^r=(r+1)^{-1}N^{r+1}(1+O(N^{-1}))$. Applied to
the last formula for $s_{j,2qj}$, this gives
\begin{eqnarray*}
 s_{j,2qj} &\simeq& 2 (\case{1}{3}\gamma\lP^2)^{-\frac{1}{4}} j^{-3}
 \left( \case{4}{11} ((q+1)j)^{\frac{11}{4}}-
 \case{4}{7}qj((q+1)j)^{\frac{7}{4}}\right.\\
 && -\left.
 \case{4}{11}((q-1)j-1)^{\frac{11}{4}}+
 \case{4}{7}qj((q-1)j-1)^{\frac{7}{4}}\right) \\
 &\simeq& \case{8}{77} (\case{1}{3}\gamma\lP^2qj)^{-\frac{1}{4}}
 q^{1/4} \left(7\left((q+1)^{\frac{11}{4}}-(q-1)^{\frac{11}{4}}\right)-
 11q\left((q+1)^{\frac{7}{4}}- (q-1)^{\frac{7}{4}}\right) \right)\,.
\end{eqnarray*}

For $q<1$ we have to split the sum in (\ref{sjn}) depending on the
sign of $qj-k$:
\begin{eqnarray*}
 s_{j,2qj} &\simeq& 2j^{-3}(\case{1}{3}\gamma\lP^2)^{-\frac{1}{4}}
 \left(\sum_{k=1}^j k
 (qj+k)^{\frac{3}{4}}- \sum_{k=1}^{qj} k(qj-k)^{\frac{3}{4}}-
 \sum_{k=qj}^j k(k-qj)^{\frac{3}{4}}\right) \\
 &=& 2j^{-3}(\case{1}{3}\gamma\lP^2)^{-\frac{1}{4}} \left(\sum_{k=1}^j \left(
 (qj+k)^{\frac{7}{4}}-qj(qj+k)^{\frac{3}{4}}\right)\right.\\
 &&+ \left.\sum_{k=1}^{qj}
 \left((qj-k)^{\frac{7}{4}}-qj(qj-k)^{\frac{3}{4}}\right)- \sum_{k=qj}^j
 \left((k-qj)^{\frac{7}{4}}+qj(k-qj)^{\frac{3}{4}}\right)  \right)\\
 &=& 2j^{-3}(\case{1}{3}\gamma\lP^2)^{-\frac{1}{4}}
 \left(\sum_{l=qj+1}^{(q+1)j} (l^{\frac{7}{4}}-qjl^{\frac{3}{4}})+
 \sum_{l=0}^{qj-1} (l^{\frac{7}{4}}-qjl^{\frac{3}{4}})
 - \sum_{l=0}^{(1-q)j} (l^{\frac{7}{4}}+qjl^{\frac{3}{4}}) \right)\\
 &=& 2j^{-3}(\case{1}{3}\gamma\lP^2)^{-\frac{1}{4}} \left(\sum_{l=1}^{(1+q)j}
 (l^{\frac{7}{4}}-qjl^{\frac{3}{4}})-
 \sum_{l=1}^{(1-q)j}(l^{\frac{7}{4}}+qjl^{\frac{3}{4}})\right)\\
 &\simeq& \case{8}{77} (\case{1}{3}\gamma\lP^2qj)^{-\frac{1}{4}}
 q^{1/4} \left( 7\left((1+q)^{\frac{11}{4}}-
 (1-q)^{\frac{11}{4}}\right)- 11q\left((1+q)^{\frac{7}{4}}+
 (1-q)^{\frac{7}{4}}\right) \right)
\end{eqnarray*}

\begin{figure}[ht]
\begin{center}
 \includegraphics[width=12cm,height=8cm,keepaspectratio]{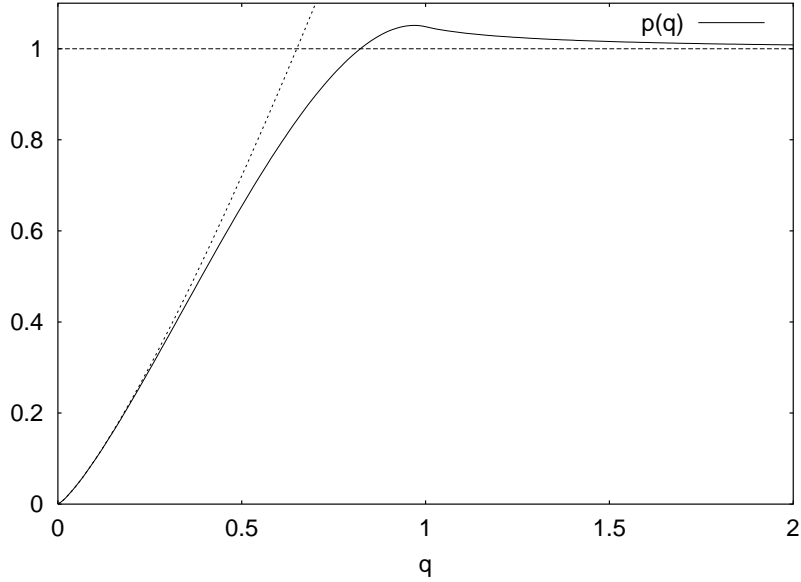}
\end{center}
\caption{The profile of $V_{qj}^{\frac{1}{6}} s_{j,2qj}$ as a function
of $q$ (which is valid for large values of $j$) compared to the
classical expectation one (dashed line); compare with
Fig.~\ref{Aprod}. The dotted line is the small-$q$ approximation
(\ref{psmall}).}
\label{Prof}
\end{figure}

Thus, the profile
\begin{equation}
 p(q):=V_{qj}^{\frac{1}{6}} s_{j,2qj}\simeq \case{8}{77}
 q^{\frac{1}{4}} \left(
 7\left((q+1)^{\frac{11}{4}}- |q-1|^{\frac{11}{4}}\right)- 11q\left(
 (q+1)^{\frac{7}{4}}- \sgn(q-1)|q-1|^{\frac{7}{4}}\right) \right)
\end{equation}
which is expected to be one classically, is independent of $j$ for
large $j$ (see Fig.~\ref{Prof}).

In the limit $q\ll1$ (but still $qj\gg1$) we have
\begin{equation}\label{psmall}
 p(q)\sim \case{12}{7} q^{\frac{5}{4}}(1-q+O(q^2))
\end{equation}
which shows that the eigenvalues $s_{j,n}$ of the inverse square root
of the scale factor increase as
\begin{equation}\label{sjnsmall}
 |s_{j,n}|=V_{\frac{1}{2}|n|}^{-\frac{1}{6}}p(\case{1}{2}|n|/j) \sim
 \case{12}{7}(\case{1}{6}\gamma\lP^2|n|)^{-\frac{1}{4}}
 \left(\case{1}{2}|n|/j\right)^{\frac{5}{4}}=
 \case{6}{7}(\case{1}{3}\gamma\lP^2j)^{-\frac{1}{4}}|n|/j\sim
 \case{12}{7}(\case{1}{3}\gamma\lP^2j)^{-\frac{5}{4}} a^2
\end{equation}
for $1\ll n\ll j$.  For $q\gg1$ the profile approaches one as
\[
 p(q)\sim 1+\case{1}{32}q^{-2}+O(q^{-4})
\]
in accordance with (\ref{sjnexpand}) for large $j$ if we use $n=2qj$.
In between these two regimes the profile attains its maximum around
the value $q=1$. It is characteristic that the value one is not
approached directly there, but only later after reaching higher
values. In fact, the maximum is at $q_{\rm max}\simeq0.97$ with value
$p(q_{\rm max})\simeq 1.05$ or more than $5\%$ above the classical
expectation, and the profile falls within $1\%$ of classical behavior
around $q=2$. This means that deviations from classical geometry are
visible there and $n$ has to be well above $2j$ for corrections to be
small.

If we have a more complicated operator containing several commutators
of holonomies with the volume operator, the same techniques can be
used. For instance, the operator $\tr(h_2[h_2^{-1},\hat{V}]
h_3[h_3^{-1},\hat{V}])$ can be written as
\[
 \tr\left( e^{-\frac{\pi}{2}\tau_1}h_3[h_3^{-1},\hat{V}]
 e^{\frac{\pi}{2}\tau_1} h_3[h_3^{-1},\hat{V}]\right)= \case{1}{2}
 \tr\left( (1-2\tau_1) h_3[h_3^{-1},\hat{V}]
 (1+2\tau_1) h_3[h_3^{-1},\hat{V}]\right)
\]
if we choose the gauge in which $h_I=\exp(c\tau_I)$ (which does not
change the gauge invariant operator). Here, we can use the basic
commutators (\ref{commj}) with an arbitrary representation $j$, and
after taking the trace we obtain an operator which is diagonal in the
$|n\rangle$-basis. Regarding the effect of changing $j$, such an
operator will have properties similar to those of the simple one
studied in more detail before. In particular, semi-classical behavior
will set in if $n$ is well above $2j$. One important operator of this
kind is the Hamiltonian constraint; thus also the dynamics will be
close to the semi-classical one only for large volume with $n\gg 2j$.

\section{Conclusions}
\label{Concl}

The calculations of this paper have shown that there is a significant
freedom in choosing particular expressions for a quantization of an
inverse power of the volume in isotropic quantum geometry. {\em Yet,
important qualitative aspects are unaltered by these modifications.}
In particular, all operators studied here annihilate the state
$|0\rangle$ which corresponds to the classical singularity. This fact
has been crucial in the proof of a non-singular cosmological evolution
\cite{Sing,IsoCosmo}, so these results are insensitive to the
quantization ambiguities studied here. Technically, the explanation
for this phenomenon is as follows (see also \cite{InvScale}). Quantum
geometry provides a natural procedure \cite{QSDI,QSDV} to quantize
classically divergent quantities like the inverse volume resulting in
bounded operators: writing co-triad components as Poisson brackets
between holonomies and the volume allows one to absorb inverse powers
of the volume. Since the Poisson bracket acts as a derivative on
$V=\sqrt{|\det (E^a_i)|}$ with respect to triad components, there will
always be a factor of the sign of $\det (E^a_i)$ which is defined to
be zero for degenerate triads (the sign appears explicitly in our
definitions of $r_k$ and $s_j$ in (\ref{rkdef}), (\ref{sjdef})). This
explains why quantizations of inverse powers of the volume should
always annihilate the degenerate state $|n\rangle$, a fact which has
been confirmed here for several classes of different
quantizations. Note that the presence of the sign is a direct
consequence of the techniques which lead to well-defined operators; it
has no effect in the classical limit since the classical description
requires non-degenerate triads.

Whereas in the first family of ambiguities (Section \ref{AmTriad}) the
suppression of the classical divergence was located at only a few
states close to the classical singularity (Fig.~\ref{Vinv}), in the
second family it took place in a large range (Section \ref{AmHol})
depending on the parameter $j$ (Fig.~\ref{Ainv}). An immediate
consequence is that $j$ cannot be arbitrarily large if we want to have
classical behavior at observable scales. While this still leaves a
large freedom in the possible values, it severely limits the
applicability of one conclusion of \cite{Gaul}. In that paper, the
idea was to {\em use\/} the ambiguity to find a Hamiltonian obeying a
`crossing symmetry' \cite{SurfaceSum}, but it was shown that this is
not possible for any finite value of $j$. The only open possibility is
to take an infinite linear combination of Hamiltonian constraint
operators with different $j$, reaching arbitrarily large values. The
results of the present paper imply that this would spoil the classical
limit of the theory unless the coefficients in the series which
defines the constraint drop off very fast for large $j$. While it is
true that any term for {\em fixed\/} $j$ in the series has the correct
classical limit at large volume, i.e.\ if $n\gg j$, unfortunately we
have to perform the $j\to\infty$ limit first if we define the
constraint as an infinite series, and we can never fulfill the
condition $n\gg j$.

Choosing large values for the parameters of the quantization
ambiguities can lead to observable effects. To see this, we take the
operators for inverse powers of the isotropic volume as models for
more complicated expressions in the full theory (most notably, matter
Hamiltonians which all involve the quantization techniques employed
here). Instead of the total volume in our formulas, we would have the
relevant scale of the physical process under investigation. (Another
aspect which is not considered here is that semiclassical states
should be used which are peaked in both the connection and the triad;
see \cite{GCSI,CohState,FockGrav} for proposals. Here we are working
with eigenstates of the triad which give an unsharp connection, but we
can interpret the present results as lower bounds for corrections
since states which do not have sharp values for the triad would have
an additional uncertainty which can only increase correction terms.)
Correction terms to classical behavior are obtained by expanding
expectation values in a parameter which is given by the Planck length
divided by the relevant length scale. Qualitative calculations for
quantum gravity corrections in matter Hamiltonians have been done in
\cite{GRB,Correct1,Correct2} where the coefficients in an expansion
have been assumed to be of order one (see also
\cite{QFTonCSTI,QFTonCSTII} for a proposed derivation from the full
theory which goes beyond a purely phenomenological analysis). If this
is the case, one can hope to find observational evidence for
deviations from the classical structure of space-time only if the
corrections are of first order in the Planck length. However, if one
does not introduce parity violating terms into the semiclassical
states used for the calculation \cite{GRB}, the corrections start
at second or fourth order \cite{Correct2}, depending on the assumed
properties of semiclassical states, the latter of which coincides with
our formulas (\ref{rkexpand}) and (\ref{sjnexpand}) (one can see that
the high order of those corrections is related to the conservation of
parity: $r_{k,n}$ in (\ref{rkn}) and $s_{j,n}$ in (\ref{sjn}) are
either even or odd in $n$ and so their absolute values are parity
invariant; this implies that expansions of these formulas in $n^{-1}$
can only contain even powers, which translates to corrections at least
of the order $\lP^4$). On the other hand, the results of the present
paper demonstrate that the magnitude of coefficients in the correction
terms can depend significantly on quantization ambiguities. Therefore,
before one can make precise predictions one must find a way to fix the
ambiguities. As the expansions (\ref{rkexpand}) and (\ref{sjnexpand})
show, the fundamental scale for correction terms is effectively given
by $k^{\frac{1}{4}}\sqrt{\gamma}\lP$ and $\sqrt{j\gamma}\lP$,
respectively, rather than just by $\sqrt{\gamma}\lP$. The value of
$\gamma$ is fixed by computations of black hole entropy
\cite{ABCK:LoopEntro,IHEntro} and smaller than one, so the only free
parameters are the ambiguity parameters $k$ and $j$ (in the families
considered here). Choosing large values for these parameters will lift
the effective fundamental scale and thereby increase correction
terms. Note that this mechanism does not affect the Planck scale which
still controls the discreteness of space. But $k$ and $j$ control the
set-in of deviations from classical behavior, and by choosing large
values we can have corrections to classical behavior at arbitrary
scales even in regions where the discreteness of space is
insignificant. However, one has to expect additional constraints on
the parameters since large values also modify the formulas at small
scales: In the $k$-family, large values of the parameter blow up
curvatures at small $n$ (recall that $r_{k,2}$ increases exponentially
in $k$), whereas the $j$-family leads to a behavior (\ref{sjnsmall})
for $n$ smaller than $j$, which is very different from the inverse
square root of the scale factor which is approached at large volume.

It may be surprising that a quantization ambiguity in quantum gravity
can have observable effects at ordinary scales rather than just at the
Planck scale. However, the classical limit $\hbar\to0$ always includes
a large volume limit $n\to\infty$ because the relation
$V\sim(\case{1}{6}\gamma\lP^2|n|)^{\frac{3}{2}}$ would imply zero
volume otherwise \cite{SemiClass}. Therefore, exact classical behavior
can only be expected in this limit in which we have $n\gg j$ for any
fixed $j$. In particular, at observable scales the classical limit
would not have been performed completely and so quantum effects there
are less surprising.

So far, we have only discussed kinematical aspects. Since
quantizations of the Hamiltonian constraint \cite{QSDI,cosmoIII} also
contain the characteristic commutator, they have the same
ambiguities. Ambiguities in the commutator itself affect the
coefficients in the constraint equation in a way as discussed
previously. But the constraint also contains additional holonomies
which quantize components of the curvature of $A_a^i$ and do not
appear in the form of commutators with the volume operator. This part
of the constraint generates the structure of the discrete time
evolution equation as a difference equation \cite{cosmoIV,IsoCosmo},
whereas the commutator only affects the coefficients appearing in the
equation. For the holonomy part we only have the second ambiguity by
choosing higher representations, which we studied in Section
\ref{AmHol}. If we use a representation (\ref{hmatrix}) with spin $j$
for all holonomies, each one will contribute factors $e^{ikc}$ with
$k$ between $-j$ and $j$ acting as multiplication operators. In flat
models, we have four holonomies, so the highest power occuring in a
multiplication operator is $e^{4ijc}$ which maps a state $|n\rangle$
as in (\ref{n}) to $|n+8j\rangle$. After transforming to the triad
representation \cite{IsoCosmo}, this leads to a difference equation of
order $16j$ (reaching from $n-8j$ to $n+8j$; this is only the counting
for the Euclidean part, the order in the Lorentzian constraint will be
twice as large). Thus, choosing a large $j$ implies a large number of
independent solutions. Even though most of them ($16j-2$) are
non-pre-classical (i.e.\ changing significantly at the Planck scale,
see \cite{DynIn}) and have to be excluded for a semiclassical
analysis, a small number of those solutions seems preferable which
would allow only small values for $j$. Therefore, a better
understanding of the status of non-pre-classical solutions could
reduce the freedom in quantization ambiguities.

Note, however, that the ambiguity parameters $j$ in the gravitational
and the matter Hamiltonian need not be identical. In particular, the
matter Hamiltonian can have a large $j$ even if one is forced (or
prefers) to use a small $j$ for the gravitational part. The matter
Hamiltonian then has a significantly different dependence on the scale
factor $a$ at small volume which generically leads to inflation
\cite{Inflation}.

\section*{Acknowledgements}

The author is grateful to A.\ Ashtekar, M.\ Gaul, H.\
Morales-T\'ecotl, H.\ Sahlmann, and T.\ Thiemann for discussions.
This work was supported in part by NSF grant PHY00-90091 and the
Eberly research funds of Penn State.

\end{document}